\documentclass[11pt,a4paper]{article}

\usepackage{amssymb}

\usepackage[dvips]{graphicx}

\begin{document}

\title{\bf Structure of three-loop contributions to the $\beta$-function of
${\cal N}=1$ SQED with $N_f$ flavors, regularized by the dimensional reduction.}

\author{
S.S.Aleshin,\\
{\small{\em Moscow State University,}}\\
{\small{\em physical faculty, department of theoretical physics,}}\\
{\small{\em 119991, Moscow, Russia}},\\
\\
A.L.Kataev,\\
{\small{\em Institute for Nuclear Research of the Russian Academy of Science,}}\\
{\small {\em 117312, Moscow, Russia}},\\
\\
K.V.Stepanyantz\\
{\small{\em Moscow State University,}}\\
{\small{\em physical faculty, department of theoretical physics,}}\\
{\small{\em 119991, Moscow, Russia}}}

\maketitle

\begin{abstract}
In the case of using the higher derivative regularization for $N=1$ SQED with $N_f$ flavors the loop integrals giving the $\beta$-function are integrals of double total derivatives in the momentum space. This feature allows to reduce one of the loop integrals to an integral of the $\delta$-function and to derive the NSVZ relation for the renormalization group functions defined in terms of the bare coupling constant. In this paper we consider $N=1$ SQED with $N_f$ flavors regularized by the dimensional reduction in the $\overline{\mbox{DR}}$-scheme. Evaluating the scheme-dependent three-loop contribution to the $\beta$-function proportional to $(N_f)^2$ we find the structures analogous to integrals of the $\delta$-singularities. After adding the scheme-independent terms proportional to $(N_f)^1$ we obtain the known result for the three-loop $\beta$-function.
\end{abstract}

\unitlength=1cm


\vspace*{-18.0cm}

\begin{flushright}
INR-TH-2015-028
\end{flushright}

\vspace*{17.0cm}

%%%%%%%%%%%%%%%%%%%%%%%%%%%%%%%%%%%

\section{Introduction}
\hspace{\parindent}

The renormalization group (RG) functions of the ${\cal N}=1$ supersymmetric quantum electrodynamics (SQED) with $N_f$ flavors are known to be related by the equation

\begin{eqnarray}\label{NSVZ}
\beta(\alpha_{0})=\frac{\alpha^{2}_{0}N_{f}}{\pi}\Big(1-\gamma(\alpha_{0})\Big),
\end{eqnarray}

\noindent
which is called ``the exact NSVZ $\beta$-function'' \cite{Novikov:1983uc,Jones:1983ip,Vainshtein:1986ja,Novikov:1985rd,Shifman:1986zi,Shifman:1985fi}. It was originally obtained for the RG functions defined in terms of the bare coupling constant \cite{Novikov:1983uc,Vainshtein:1986ja,Novikov:1985rd,Shifman:1985fi}. A rigorous derivation of this relation by summing supergraphs was also given in \cite{Stepanyantz:2011jy,Stepanyantz:2014ima} for the RG functions defined in terms of the bare coupling constant using the higher derivative (HD) regularization \cite{Slavnov:1971aw,Slavnov:1972sq} generalized to the supersymmetric case \cite{Krivoshchekov:1978xg,West:1985jx}. This derivation is based on the observation that all loop integrals giving the $\beta$-function are integrals of double total derivatives in the momentum space. The factorization into integrals of total derivatives was first noted in Ref. \cite{Soloshenko:2003nc}, and the factorization into integrals of double total derivatives was found in Ref. \cite{Smilga:2004zr}. This feature was also verified by the calculations in the lowest orders of the perturbation theory in non-Abelian supersymmetric theories \cite{Pimenov:2009hv,Stepanyantz:2011bz,Kazantsev:2014yna}, and in theories with ${\cal N}=2$ supersymmetry \cite{Buchbinder:2014wra,Buchbinder:2015eva}. The factorization of loop integrals for the Adler $D$-function of ${\cal N}=1$ SQCD into integrals of double total derivatives was also proved in all orders of the perturbation theory \cite{Shifman:2014cya,Shifman:2015doa}.

The factorization property valid for supersymmetric theories regularized by higher derivatives gives rise to the exact NSVZ $\beta$-function for the RG functions defined in terms of the bare coupling constant (at least for ${\cal N}=1$ SQED with $N_f$ flavors). Due to this feature, one of the loop integrals is an integral of the $\delta$-function and can be calculated analytically. If the RG functions are defined in terms of the renormalized coupling constant, the factorization into double total derivatives allows to construct a concrete prescription which in the Abelian case determine the NSVZ scheme in all orders in the case of using the HD regularization \cite{Kataev:2013eta,Kataev:2013csa,Kataev:2014gxa}.

Nevertheless, supersymmetric theories are mostly regularized by the dimensional reduction (DRED) \cite{Siegel:1979wq}, which is however mathematically inconsistent \cite{Siegel:1980qs} in higher orders \cite{Avdeev:1981vf,Avdeev:1982xy}. Therefore, it is interesting to investigate loop corrections to the $\beta$-function in supersymmetric theories regularized by DRED and to look for structures similar to the integrals of the $\delta$-functions, which appear in the case of using the HD regularization. The minimal order in which the $\beta$-function is scheme dependent corresponds to the three-loop approximation. (For the considered theory at the three-loop level the inconsistency of DRED is not essential, see, e.g., \cite{Mihaila:2013wma}.)

In this paper we explicitly demonstrate that structures similar to the integrals of the $\delta$-functions appear in the three-loop contributions to the $\beta$-function proportional to $(N_f)^2$. These terms are especially interesting, because they are scheme-dependent, while the terms proportional to $N_f$ do not depend on the subtraction scheme \cite{Kataev:2013csa}. Explicitly calculating the $(N_f)^2$ terms and adding the known scheme-independent contributions proportional to $(N_f)^1$ we obtain the total three-loop $\beta$-function of ${\cal N}=1$ SQED with $N_f$ flavors in the $\overline{\mbox{DR}}$-scheme. The result agrees with Ref. \cite{Jack:1996vg}, in which the structures similar to integrals of the $\delta$-functions were not discovered.

\section{${\cal N}=1$ SQED with $N_{f}$ flavors and its regularization by DRED}
\hspace{\parindent}

In this paper we consider massless ${\cal N}=1$ SQED with $N_{f}$ flavors. In terms of the ${\cal N}=1$ superfields the action of this theory has the form

\begin{eqnarray}\label{Action}
S=\frac{1}{4e_{0}^{2}}\mbox{Re}\int d^4x\,d^2\theta\, W^{a}W_{a}+
\sum_{i=1}^{N_{f}}\frac{1}{4}\int d^4x\,d^4\theta\,\Big(\phi_{i}^{*}e^{2V}\phi_{i}+\widetilde{\phi}_{i}^{*}e^{-2V}\widetilde{\phi}_{i}\Big),
\end{eqnarray}

\noindent
where $e_{0}$ is the bare coupling constant.

Quantum corrections in this theory will be calculated using the regularization by means of DRED \cite{Siegel:1979wq}. This implies that the algebra of supersymmetric covariant derivatives is the same as in four dimensions, while the integrals over loop momentums (which remain after calculating Grassmannian integrals over all $\theta$-s except for the one) are calculated in the dimension $d$. Note that the coupling constant in $d\ne 4$ dimensions is dimensionful, so that it is convenient to make the substitution

\begin{equation}
e_0^2 \to e_0^2 \Lambda^\varepsilon
\end{equation}

\noindent in the regularized theory, where $\Lambda$ is a constant with the dimension of mass and $\varepsilon \equiv 4-d$. In this case $e_0$ will be also dimensionless.

A part of the effective action corresponding to the two-point Green function can be written in the form

\begin{eqnarray}\label{Gamma_Correction}
&& \Gamma^{(2)}-S_{gf}=-\frac{1}{16\pi}\int\frac{d^{4}p}{(2\pi)^{4}}d^{4}\theta\, V(-p,\theta)\partial^{2}\Pi_{1/2}V(p,\theta)\, d^{-1}(\alpha,\mu/p)
\nonumber\\
&& +\frac{1}{4}\sum_{i=1}^{N_{f}}
\int\frac{d^{4}p}{(2\pi)^{4}}d^{4}\theta\,\Big(\phi_{i}^{*}(-p,\theta)\phi_{i}(p,\theta)+\widetilde{\phi}_{i}^{*}(-p,\theta)\widetilde{\phi}_{i}(p,\theta)\Big)\, G(\alpha,\mu/p),
\end{eqnarray}

\noindent
where $\partial^{2}\Pi_{1/2}=-D^{a}\bar {D^{2}}D_{a}/8$ denotes the supersymmetric transversal projection operator, $\alpha=\alpha(\mu)$ is the renormalized coupling constant, and $\mu$ is the renormalization scale.\footnote{Usually, within the dimensional technique one sets $\Lambda=\mu$, but in this paper we do not impose this condition in order to show explicitly that all subsequent equations are similar to the ones obtained with the HD regularization.} The transversality of quantum corrections to the two-point Green function of the gauge field follows from the Ward identity.

\section{Structure of the $(N_f)^2$ three-loop contributions to the $\beta$-function}
\hspace{\parindent}

The three-loop expression for the function $d^{-1}(\alpha_{0},\Lambda/p)$ of the considered theory can be constructed using the results of Ref. \cite{Soloshenko:2003nc}.\footnote{In Ref. \cite{Soloshenko:2003nc} the HD method was used for regularization. However, the DRED integrals can be easily derived from the ones obtained with the HD regularization.} Keeping in the three-loop approximation only scheme-dependent terms proportional to $(N_f)^2$, we can rewrite it in the form

\begin{eqnarray}\label{Explicit_D_Function}
d^{-1}(\alpha_{0},\Lambda/p)=\frac{1}{\alpha_0} + I_{1}+I_{2}+I_{3}+O(\alpha_{0}^{2}N_{f})+O(\alpha_{0}^{3}),
\end{eqnarray}

\noindent
where $\alpha_0=e_0^2/4\pi$, the integrals $I_1$ and $I_2$ are the one- and two-loop contributions to the function $d^{-1}$, and $I_3$ is a part of the three-loop contribution proportional to $(N_f)^2$. The part of the three-loop contribution proportional to $N_f$ is not considered in this paper. In the explicit form the integrals $I_i$ are written as

\begin{eqnarray}\label{Interals}
&& I_{1}=8\pi N_{f} \Lambda^{\varepsilon}\int\frac{d^{d}k}{(2\pi)^{d}} \frac{1}{k^{2}(k+p)^{2}};\vphantom{\Bigg(}\\
&& I_{2}=16\pi e_{0}^{2}N_{f} \Lambda^{2\varepsilon}\int\frac{d^{d}k}{(2\pi)^{d}}\frac{d^{d}q}{(2\pi)^{d}}\left(\frac{2}{k^{2}(k+q)^{2}q^{2}(q+p)^{2}}\right. \nonumber\\
&&\qquad\qquad\quad \left. -\frac{1}{(k+q)^{2}(k+q+p)^{2}q^{2}(q+p)^{2}}
-\frac{p^{2}}{k^{2}(k+q)^{2}(k+p+q)^{2}q^{2}(q+p)^{2}}\right);\\
&& I_{3}=-32\pi e_{0}^{4}(N_{f})^{2} \Lambda^{3\varepsilon}\int\frac{d^{d}k}{(2\pi)^{d}}\frac{d^{d}q}{(2\pi)^{d}}\frac{d^{d}t}{(2\pi)^{d}}\frac{1}{t^{2}(t+k)^{2}}
\left(\frac{2}{k^{2}(k+q)^{2}q^{2}(q+p)^{2}}\right.\nonumber\\
&&\qquad\qquad\quad \left. -\frac{1}{(k+q)^{2}(k+q+p)^{2}q^{2}(q+p)^{2}}
-\frac{p^{2}}{k^{2}(k+q)^{2}(k+p+q)^{2}q^{2}(q+p)^{2}}\right).\qquad
\end{eqnarray}

\noindent
In the case of using the HD regularization the expressions similar to $I_2$ and $I_3$ are related to the one- and two-loop contributions to the function $\ln G$, respectively, due to the factorization of the corresponding integrals into integrals of double total derivatives in the limit $p\to 0$. In the case of using the dimensional technique it is impossible to take the limit $p\to 0$, because loop integrals are proportional to $(\Lambda/p)^{n\varepsilon}$, where $n$ is an integer. That is why with DRED such a factorization into integrals of double total derivatives evidently does not take place. Nevertheless, one can try to find its analog. For this purpose, we first consider the two-loop integral $I_2$ and add to it

\begin{equation}\label{Vanishing_Total_Derivative}
0 =16\pi e_{0}^{2}N_{f} \Lambda^{2\varepsilon}\int\frac{d^{d}k}{(2\pi)^{d}}\frac{d^{d}q}{(2\pi)^{d}}\frac{\partial}{\partial q^{\mu}}\left(\frac{q^{\mu}}{q^{2}(q+p)^{2}k^{2}(q+k)^{2}} \Big(1-\frac{p^2}{2(q+k+p)^2}\Big)\right).
\end{equation}

\noindent
Then, after some transformations the result can be rewritten as

\begin{eqnarray}
&&\hspace*{-5mm} I_{2}=16\pi e_{0}^{2}N_{f} \Lambda^{2\varepsilon}
\int\frac{d^{d}k}{(2\pi)^{d}}\frac{d^{d}q}{(2\pi)^{d}}
\left(-\frac{\varepsilon}{k^{2}(k+q)^{2}q^{2}(q+p)^{2}} -2p^{\mu}\frac{\partial}{\partial p^{\mu}}\frac{1}{k^{2}(k+q)^{2}q^{2}(q+p)^{2}}\right.\nonumber\\
&&\hspace*{-5mm} +\frac{1}{2}p^{\mu}\frac{\partial}{\partial p^{\mu}}\frac{1}{(k+q)^{2}(k+q+p)^{2}q^{2}(q+p)^{2}}+\frac{1}{2}p^{2}p^{\mu}\frac{\partial}{\partial p^{\mu}}\frac{1}{k^{2}q^{2}(q+p)^{2}(q+k)^{2}(q+k+p)^{2}}\nonumber\\
&&\hspace*{-5mm}\left. +\frac{\varepsilon p^{2}}{2k^{2}q^{2}(q+p)^{2}(q+k)^{2}(q+k+p)^{2}}\right).
\end{eqnarray}

\noindent
From dimensional considerations it is easy to see that all two-loop integrals in this equation are proportional to $(\Lambda/p)^{2\varepsilon}$. This allows to make easily the differentiation with respect to the momentum $p^\mu$, after which we obtain

\begin{eqnarray}
&& I_{2}=\varepsilon I_{2}+16\pi e_{0}^{2}N_{f} \Lambda^{2\varepsilon}
\int\frac{d^{d}k}{(2\pi)^{d}}\frac{d^{d}q}{(2\pi)^{d}}
\left(\frac{\varepsilon}{k^{2}(k+q)^{2}q^{2}(q+p)^{2}}\right.
\nonumber\\
&&\qquad\qquad\qquad\qquad\qquad\quad\qquad\qquad\qquad
\left. +\frac{p^{2}(-2+\varepsilon)}{2 k^{2}q^{2}(q+p)^{2}(q+k)^{2}(q+k+p)^{2}}\right).\qquad
\end{eqnarray}

\noindent
As a consequence, the considered two-loop contribution can be presented in the form

\begin{eqnarray}\label{I2}
&& I_{2}=16\pi e_{0}^{2}N_{f} \Lambda^{2\varepsilon} \left(\frac{\varepsilon}{1-\varepsilon}
\int\frac{d^{d}k}{(2\pi)^{d}}\frac{d^{d}q}{(2\pi)^{d}}
\frac{1}{k^{2}(k+q)^{2}q^{2}(q+p)^{2}}\right.\nonumber\\
&&\qquad\qquad\qquad\qquad\left. +\frac{\varepsilon-2}{2(1-\varepsilon)}
\int\frac{d^{d}k}{(2\pi)^{d}}\frac{d^{d}q}{(2\pi)^{d}} \frac{p^2}{k^{2}q^{2}(q+p)^{2}(q+k)^{2}(q+k+p)^{2}}\right).\qquad
\end{eqnarray}

\noindent
The second term in this expression is finite and proportional to $\zeta(3)$ in the limit $\varepsilon\to 0$ \cite{Chetyrkin:1980pr}. Therefore, it does not contribute to the $\beta$-function in the $\overline{\mbox{DR}}$-scheme. The first term is divergent. Below we will demonstrate that it is related to the one-loop contribution to $\ln G$. However, before doing this, we first write the three-loop contribution $I_3$ proportional to $(N_f)^2$ in a similar way. Adding to it

\begin{eqnarray}
&& 0=-32\pi e_{0}^{4} (N_{f})^{2} \Lambda^{3\varepsilon} \int\frac{d^{d}k}{(2\pi)^{d}}\frac{d^{d}q}{(2\pi)^{d}}\frac{d^{d}t}{(2\pi)^{d}} \frac{1}{t^{2}(t+k)^{2}}\,\frac{\partial}{\partial q^{\mu}}\left(\frac{q^{\mu}}{q^{2}(q+p)^{2}k^{2}(q+k)^{2}} \right.\qquad
\nonumber\\
&&\left. -\frac{p^2 q^\mu}{2 k^2 q^2 (q+p)^2 (q+k)^2 (q+k+p)^2} \right)
\end{eqnarray}

\noindent
and making the transformations similar to the ones described above we obtain

\begin{eqnarray}\label{I3}
&& I_{3}=-32\pi e_{0}^{4}(N_{f})^{2}\Lambda^{3\varepsilon}\left(\frac{2\varepsilon}{1-3\varepsilon/2}
\int\frac{d^{d}k}{(2\pi)^{d}}\frac{d^{d}q}{(2\pi)^{d}}\frac{d^{d}t}{(2\pi)^{d}}\,
\frac{1}{t^{2}(t+k)^{2} k^{2}(k+q)^{2}q^{2}(q+p)^{2}}\right.\qquad\nonumber\\
&&\left. +\frac{\varepsilon-2}{2(1-3\varepsilon/2)}
\int\frac{d^{d}k}{(2\pi)^{d}}\frac{d^{d}q}{(2\pi)^{d}}\frac{d^{d}t}{(2\pi)^{d}}\, \frac{p^2}{t^{2}(t+k)^{2} k^{2}q^{2}(q+p)^{2}(q+k)^{2}(q+k+p)^{2}}\right).
\end{eqnarray}

\noindent
Note that deriving this expression we took into account that the three-loop integrals in the considered expression are proportional to $(\Lambda/p)^{3\varepsilon}$. Also we note that, unlike the $I_2$ case, the last term is divergent. However, it is easy to see that this divergence vanishes when one expresses the bare coupling constant in terms of the renormalized one in the expression $d^{-1}- \alpha_0^{-1}$.

Using Eqs. (\ref{I2}) and (\ref{I3}) the function $d^{-1}$  can be presented as

\begin{eqnarray}\label{D_Expression}
&& d^{-1}(\alpha_{0},\Lambda/p) =  \alpha_0^{-1} + 8\pi N_{f} \Lambda^{\varepsilon}\int\frac{d^{d}k}{(2\pi)^{d}}\frac{1}{k^{2}(k+p)^{2}}\nonumber\\
&& +64\pi^2 \alpha_0 N_{f} \Lambda^{2\varepsilon} \frac{\varepsilon}{1-\varepsilon}\int\frac{d^{d}q}{(2\pi)^{d}}\frac{1}{q^{2}(q+p)^{2}}
\int\frac{d^{d}k}{(2\pi)^{d}}\frac{1}{k^{2}(k+q)^{2}}\nonumber\\
&& -512\pi^3 \alpha_0^2 (N_{f})^{2} \Lambda^{3\varepsilon}\frac{2\varepsilon}{1-3\varepsilon/2}\int\frac{d^{d}q}{(2\pi)^{d}}\frac{1}{q^{2}(q+p)^{2}}
\int\frac{d^{d}k}{(2\pi)^{d}} \frac{1}{k^{2}(q+k)^{2}} \frac{d^{d}t}{(2\pi)^{d}}\frac{1}{t^{2}(t+k)^{2}}\qquad\nonumber\\
&& +\mbox{finite terms} + O(\alpha_0^2 N_f) + O(\alpha_0^3).\vphantom{\frac{1}{2}}
\end{eqnarray}

\noindent
In this expression we do not write explicitly terms which are finite functions of the renormalized coupling constant $\alpha=\alpha(\mu)$, related to the bare coupling constant by the equality

\begin{equation}
\frac{1}{\alpha_0} = \frac{1}{\alpha} - \frac{N_f}{\pi}\Big(\frac{1}{\varepsilon} + \ln\frac{\Lambda}{\mu} + b_1\Big) + O(\alpha^2),
\end{equation}

\noindent
where $b_1$ is a finite constant depending on the renormalization scheme.

Comparing the expression (\ref{D_Expression}) with similar terms in the logarithm of the two-point Green function of the matter superfields

\begin{eqnarray}
&&\hspace*{-5mm} \ln G = -8\pi\alpha_0 \Lambda^{\varepsilon}
\int\frac{d^{d}k}{(2\pi)^{d}}\frac{1}{k^{2}(q+k)^{2}} + 64\pi^{2}\alpha_0^{2}N_{f} \Lambda^{2\varepsilon}
\int\frac{d^{d}k}{(2\pi)^{d}} \frac{d^{d}t}{(2\pi)^{d}}  \frac{1}{k^{2}(q+k)^{2} t^{2}(t+k)^{2}}\nonumber\\
&&\hspace*{-5mm} + O\left((N_f)^0 \alpha_0^2\right) + O(\alpha_0^3),\vphantom{\frac{1}{2}}
\end{eqnarray}

\noindent
we obtain

\begin{eqnarray}\label{D_Vs_LnG}
&& d^{-1}-\alpha_{0}^{-1} = 8\pi N_{f} \Lambda^{\varepsilon}\int\frac{d^{d}q}{(2\pi)^{d}}\frac{1}{q^{2}(q+p)^{2}}\nonumber\\
&& -8\pi N_{f} \Lambda^{\varepsilon}\frac{\varepsilon}{1-\varepsilon}\int\frac{d^{d}q}{(2\pi)^{d}}\frac{1}{q^{2}(q+p)^{2}}
(\ln G)_{1-\mbox{\scriptsize loop}}\nonumber\\
&& -8\pi N_{f} \Lambda^{\varepsilon}\frac{2\varepsilon}{1-3\varepsilon/2}\int\frac{d^{d}q}{(2\pi)^{d}}\frac{1}{q^{2}(q+p)^{2}}
(\ln G)_{2-\mbox{\scriptsize loops}, N_f}\nonumber\\
&& +\mbox{finite terms} + O(N_f \alpha_0^2) + O(\alpha_0^3).\vphantom{\frac{1}{q^2}}
\end{eqnarray}

The relation between the functions $d^{-1}-\alpha_0^{-1}$ and $\ln G$ was earlier found in all orders in the case of using the HD regularization \cite{Stepanyantz:2011jy,Stepanyantz:2014ima}. It is given by the equality

\begin{equation}\label{D_Vs_LnG_HD}
\frac{d}{d\ln\Lambda} \Big(d^{-1}-\alpha_{0}^{-1}\Big)\Big|_{p=0} = \frac{d}{d\ln\Lambda}\Big(\mbox{One-loop} - 16\pi^3 N_f \int \frac{d^4q}{(2\pi)^4} \delta^4(q) \ln G\Big).
\end{equation}

\noindent
where $\Lambda$ is now the dimensionful parameter in the HD term, $\mbox{One-loop}$ denotes the one-loop contribution, and the differentiation is made at a fixed value of the renormalized coupling constant. $\delta^4(q)$ appears from the integrals of double total derivatives due to the identity

\begin{equation}
\frac{\partial}{\partial q^\mu} \frac{\partial}{\partial q^\mu} \frac{1}{q^2} = -4\pi^2 \delta^4(q).
\end{equation}

\noindent
Due to the presence of $\ln G$, Eq. (\ref{D_Vs_LnG}) is an analog of Eq. (\ref{D_Vs_LnG_HD}) if the theory is regularized by DRED. Comparing these equations we conclude that in the considered integrals the expression

\begin{equation}
\frac{1}{2\pi^2} \Lambda^\varepsilon\cdot \frac{(L-1)\varepsilon}{1-L\varepsilon/2} \int \frac{d^dq}{(2\pi)^d} \frac{1}{q^2 (q+p)^2},
\end{equation}

\noindent where $L$ is a number of loops, is an analog of the structure

\begin{equation}
\int \frac{d^4q}{(2\pi)^4} \delta^4(q),
\end{equation}

\noindent
which appears after calculating integrals of double total derivatives with the HD regularization in the limit of the vanishing external momentum.

Starting from Eq. (\ref{D_Vs_LnG}) it is possible to find the expression for the $\beta$-function in the considered approximation in the $\overline{\mbox{DR}}$-scheme. The finite terms are not essential in this scheme. In our conventions, the $\overline{\mbox{DR}}$-scheme is obtained by subtracting $\varepsilon$-poles and powers of $\ln \bar\Lambda/\mu$, where $\bar\Lambda \equiv \Lambda \exp(-\gamma/2)\sqrt{4\pi}$. Calculating integrals in Eq. (\ref{D_Vs_LnG}) using the standard dimensional technique and making the renormalization we obtain that the corresponding contribution to the $\beta$-function of ${\cal N}=1$ SQED with $N_f$ flavors is

\begin{equation}
\beta_{\overline{\mbox{\scriptsize DR}}}(\alpha) = \left.\frac{d\alpha_{\overline{\mbox{\scriptsize DR}}}}{d\ln\mu}\right|_{\alpha_0=\mbox{\scriptsize const}} = \frac{\alpha^2 N_f}{\pi}\Big(1 + \frac{\alpha}{\pi} - \frac{3 N_f\alpha^2}{4\pi^2} \Big) + O(N_f \alpha^4) + O(\alpha^5).
\end{equation}

\noindent Adding the scheme-independent three-loop contributions proportional to $N_f$ \cite{Jack:1996vg} we find that the total three-loop $\beta$-function of the considered theory has the form

\begin{equation}
\beta_{\overline{\mbox{\scriptsize DR}}}(\alpha) = \frac{\alpha^2 N_f}{\pi}\Big(1 + \frac{\alpha}{\pi} - \frac{(2+3 N_f)\alpha^2}{4\pi^2} \Big) + O(\alpha^5).
\end{equation}

\noindent
One can see that it agrees with the results of Ref. \cite{Jack:1996vg}. Note that this expression for the RG function does not satisfy the NSVZ relation, which can be obtained from the $\overline{\mbox{DR}}$-result only after a special finite renormalization \cite{Jack:1996vg,Jack:1996cn}.

\section{Conclusion}
\hspace{\parindent}

We have demonstrated that the factorization of integrals giving the $\beta$-function of $N=1$ SQED with $N_{f}$ flavors regularized by higher derivatives into integrals of $\delta$-singularities has an analog in the case of using DRED at least for the three-loop scheme-dependent terms proportional to $(N_f)^2$. Instead of the integrals of $\delta$-singularities (which appear in calculating integrals of double total derivatives) a certain typical structure appears. Presumably, the existence of such a structure can allow constructing a prescription which gives the NSVZ scheme in all orders of the perturbation theory if supersymmetric theories are regularized by DRED. If it is possible, such a prescription will be an analog of the one constructed in Ref. \cite{Kataev:2013eta} for the case of using the HD regularization.

\section*{Acknowledgments}
\hspace{\parindent}

The authors are very grateful to A.O.Barvinsky for the valuable discussions.

The work of A.K. was supported by the RFBR grant No. 14-01-00647 and also by the grant for the Leading Scientific Schools NS-2835.2014.2. The work of K.S. was supported by the RFBR grant No. 14-01-00695.

\end{document}